\documentclass[aps,pra,twocolumn]{revtex4-2} \usepackage{amsmath} 
\usepackage{amssymb} 
\usepackage{graphicx} 
\begin{document} 
\title{Pressure gradient-driven plasma flows and magnetogenesis} 
\author{Z. H. Saleem} 
\affiliation{Argonne National Laboratory, 9700 S. Cass Ave, Lemont, IL 60439, USA} 
\author{H. Saleem} 
\email{saleemhpk@hotmail.com} 
\affiliation{Theoretical Research Institute, Pakistan Academy of Sciences, 3-Constitution Avenue, G-5/2, Islamabad 44000, Pakistan} 
\affiliation{Department of Physics, School of Natural Sciences (SNS), National University of Sciences and Technology (NUST), H-12, Islamabad 44000, Pakistan} \affiliation{Department of Space Science, Institute of Space Technology (IST), Islamabad 44000, Pakistan} 

\begin{abstract} We present a self-consistent two-fluid theory demonstrating that pressure gradients simultaneously generate plasma flows and magnetic fields. We show that compatibility between ion momentum balance and mass conservation imposes a previously unrecognized constraint on plasma evolution: the total pressure must satisfy the Laplace equation, $\nabla^2 p = 0$. This condition yields a class of exact analytical solutions in which pressure-driven flows and Biermann-type magnetic fields emerge together. Application of the model to a galactic gas clump reveals that, under thermal pressure, electrons and ions move almost together, giving rise to weak currents and consequently very small seed magnetic fields. Ion dynamics are also important for determining the seed magnetic-field generation time $\tau_B$ and for estimating the ion flow velocity. The model is further applied to laser-produced plasma to describe its short-time evolution. The present theory provides a unified, self-consistent description of pressure-driven flow generation and magnetogenesis in both astrophysical and laboratory plasmas. \end{abstract} 

\maketitle

\section{Introduction}
Magnetic fields of order microgauss have been observed in nearly all galaxies, and it is widely believed that extremely weak seed fields were generated during galaxy formation time which were subsequently amplified through dynamo processes \citep{fitt1993magnetic,widrow2002origin}. The origin of such seed magnetic fields was first investigated in the 1970s \citep{harrison1970generation,harrison1973origin}. Since then, galactic and extragalactic magnetic fields have remained an active area of research, attracting considerable attention over both past and recent decades \citep{kronberg1994extragalactic,zweibel1997magnetic,widrow2002origin,de2025magnetic,nazareth2025galactic}. A theoretical mechanism for magnetic-field generation in rotating stars, arising from non-parallel gradients of electron density and temperature, was proposed by \citet{biermann1950ursprung}. The magnitude of seed magnetic field generated in a galactic gas clump was estimated by employing the Biermann mechanism assuming the electron pressure to be time-independent \citet{lazarian1992diffusion}. 

The observation of magnetic fields of the order of a megagauss in early classical laser-plasma experiments \citep{stamper1971spontaneous,stamper1975faraday,raven19780} stimulated extensive interest within the plasma physics community and motivated investigations into a variety of magnetic-field generation mechanisms \citep{al1974contribution,tidman1976magnetic,jones1983magnetic,bol1983spontaneous,haines1997saturation}. The magnetic field generated in a classical laser-produced plasma was also estimated using Biermann’s theory \citep{brueckner1974laser}. In Biermann battery mechanism, the ion dynamics are ignored and electrons are assumed to be inertialess assuming characteristic time scale for magnetic field generation to be much shorter than the ion plasma oscillation time and much longer than the electron plasma oscillation time. However, a few years ago  \citep{saleem20223d,saleem2024generation}, analytical solutions of the two-fluid plasma equations were obtained to explain the simultaneous generation of plasma flow and magnetic field, as well as the formation of spicules in the solar atmosphere. In almost all of the above mentioned models, the density and temperature profiles were chosen independently and therefore they lacked full self-consistency. 

In the present work, we demonstrate that mainly the flows of electrons and ions are generated by inhomogeneous plasma pressure which in turn give rise to weak seed magnetic fields in galactic gas clumps. The thermal force moves the electrons and ions with comparable velocities and hence the currents are small. The compatibility between ion momentum balance and mass conservation imposes a fundamental structural constraint on the system: the total pressure, $p=n(T_e+T_i)$ must satisfy the Laplace equation, $\nabla^2 p=0$. Consequently, a class of solutions characterized by harmonic pressure distribution naturally emerges, enabling the simultaneous generation of plasma flows and magnetic fields.


\section{Theoretical Model}
The seed magnetic field generation time $\tau_B$ is assumed to be much longer than electron plasma oscillation time $t_e=\omega_{pe}^{-1}$ where $\omega_{pe}=(\frac{4 \pi n_e e^2}{m_e})^{1/2}$ is the electron plasma oscillation frequency viz $t_e \ll \tau_B$. In the limit $m_e \rightarrow 0$, the electron momentum conservation equation becomes,
\begin{equation}
	0 = - e n \left(\mathbf{E} + \frac{\mathbf{v}_e \times \mathbf{B}}{c}\right) - \nabla p_e
\end{equation}
The small effect of weak generated magnetic field can be ignored to get \citep{biermann1950ursprung}, 
\begin{equation}
	- e n \mathbf{E} - \nabla p_e =0
\end{equation}
The curl of above equation along with Faraday's law yields,
\begin{equation}
	\partial_t \mathbf{B} = - (c/e) (\frac{\nabla n_e}{n_e} \times \nabla T_e) 
\end{equation}
Equation (3) describes the Biermann battery mechanism \citep{biermann1950ursprung}. 
For ions, the equation of motion in standard form is,
\begin{equation}
m_i n_i (\partial_t + \mathbf{v}_i \cdot \nabla)\mathbf{v}_i = \phantom{-} e n_i \left(\mathbf{E} + \frac{\mathbf{v}_i \times \mathbf{B}}{c}\right) - \nabla p_i
\end{equation}
Then the mass conservation equation for ions becomes,
\begin{equation}
	\partial_t n + \nabla \cdot (n \mathbf{v}_i) = 0   
\end{equation} 
where we have used quasi-neutrality $n_e \simeq n_i = n$.
Amperes' law relates electron velocity with ion velocity,
\begin{equation}
{\bf v}_e = {\bf v}_i - \frac{c}{4 \pi n e} (\nabla \times {\bf B})
\end{equation}
If $\mid (\mathbf{v}_i \cdot \nabla)\mathbf{v}_i \mid \ll \mid \partial_t {\bf v}_i \mid $ in Eq. (4), it reduces to a simpler form,
\begin{equation}\label{eq:ionlinear}
m_i \partial_t \mathbf{v}_i = - \nabla p / n 
\end{equation}
where $p = n(T_e + T_i)$. This simplification is valid only if the generation time $t=\tau_B$ fulfills the following condition,
\begin{equation}\label{eq:taubcondition}
	 \tau_B \ll L/v_i
\end{equation}
where $L$ is the characteristic gradient scale length. Magnetic back-reaction is likewise small because Ampère’s law gives  $|\mathbf{v}_i - \mathbf{v}_e| \sim (c/4\pi e n)(B/L) =\frac{\lambda_i}{L}(\lambda_i \Omega_i) \ll 1$ where $\lambda_i = c/\omega_{pi}$ is the ion skin depth, $c_s = (\frac{T_e}{m_i})^{1/2}$ and $\Omega_i= (\frac{e B}{m_i c})$. 



\begin{figure}[t]
\centering
\includegraphics[width=\columnwidth]{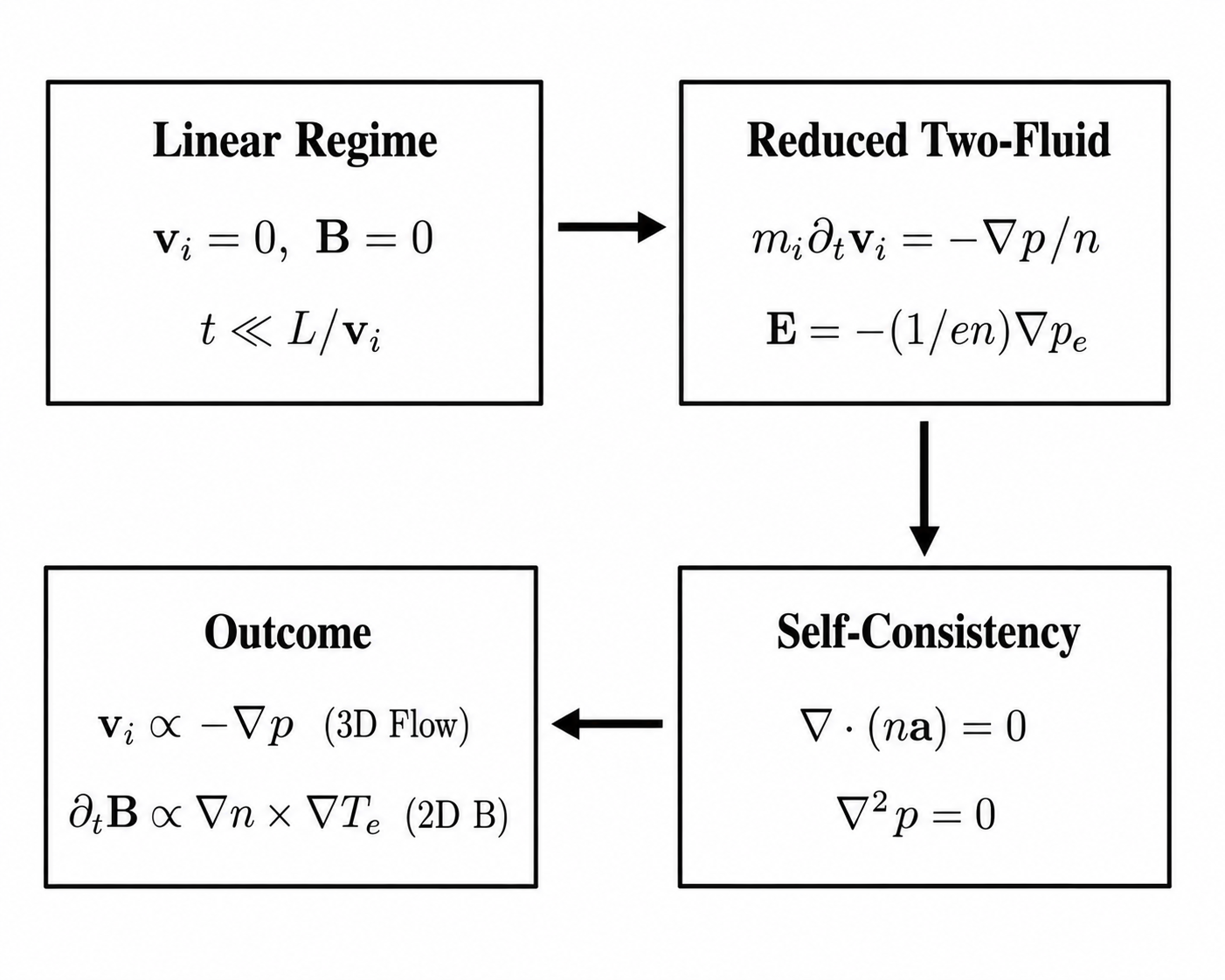}
\caption{
The schematic highlights how pressure acts as the organizing principle linking flow generation and magnetogenesis in linear limit.
}
\label{fig:schematic}
\end{figure}

\begin{figure}[t]
\centering
\includegraphics[width=\columnwidth]{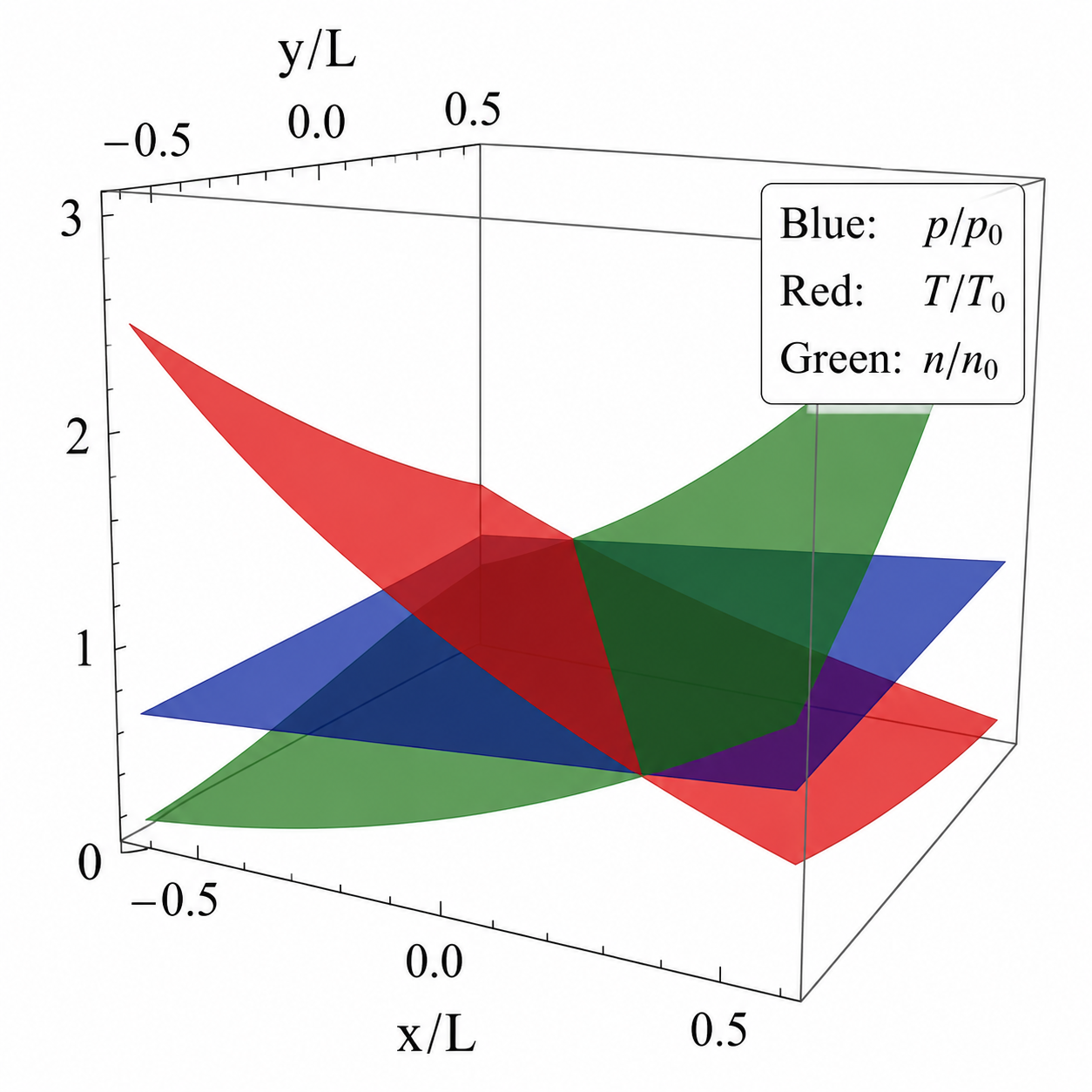}
\caption{
Normalized profiles of pressure $p/p_0$, temperature $T/T_0$, and density $n/n_0$ for the linear harmonic-pressure solution, shown over the $(x,y)$ plane at fixed $z=0$. Here $p_0$, $n_0$ and $T_0$ are the values at $(x,y,z)=(0,0,0)$. 
}
\label{fig:linear_3d_profiles}
\end{figure}


\section{Harmonic Pressure Constraint:}
If $\partial_t n = 0$, then ion continuity equation demands  $\nabla \cdot (n \mathbf{v}_i) = 0$ which imposes a structural constraint on spatial profile of total pressure in Eq. (7),
\begin{equation}
	\nabla^2 p = 0
\end{equation}
The simplest form of $p$ can be,
\begin{equation}
 p(x,y,z) = p_0 (1+ \alpha x + \beta y + \gamma z)   
\end{equation}
where $\alpha, \beta, \gamma$ are constants and $p_0= p(0,0,0)$.
If the temperature is assumed to obey the following structure,
\begin{equation}
T= T_0 e^{\alpha x+ \beta y}
\end{equation}
then density will take the form,
\begin{equation}
n(x,y,z) = n_0 e^{-\sigma} (1+ \alpha x + \beta y + \gamma z) 
\end{equation}
where $\sigma= \alpha x+ \beta y$. More structured magnetic fields can be produced by modulated pressure profiles.
There are several possible exact analytical solutions with harmonic pressure which generate plasma flow and magnetic field simultaneously. For example, we may choose the pressure and temperature as $p (x,y,z) = p_0 e^{\mu z} \cos (\alpha x+ \beta y)$ and $T= T_0 e^{\mu z}$ where $\mu$ is a constant.
Then density has to take the form $n (x,y) = n_0 \cos (\alpha x + \beta y)$,
This solution imposes an additional restriction $\mu^2 = \alpha^2 + \beta^2$ on the constants. 


\section{Astrophysical Plasma:}
In previous investigation \citep{lazarian1992diffusion}, the electron temperature at the border of the galactic gas clump was taken to be $T_{out}=10^{6}\,$K while inside the cloud, temperature was much smaller i.e. $T_{in}\ll T_{out}$, where $T_{in}=T_{e0}$. Then $\nabla T_e \simeq 10^{6}/R$, where $R=10^{4}\,$pc is the cloud radius and one parsec (pc) $=3\times10^{18}\,$cm. The electron density profile was assumed to be exponential, viz. $\nabla n_e/n_e = 1/h$, where $h=10^{2}\,$pc is the thickness of the cloud. Using Eq.~(3), one obtains the space-independent seed magnetic field of order $|{\bf B}|\simeq 3\times10^{-17}\,$G.

We now use the present model and choose the density and temperature profiles to be as close as possible to those of Lazarian to highlight the role of ion dynamics. For this purpose, let $p=p_{e0}(1+\beta y)$ and $n_e=n_{e0}e^{\alpha z}$, which give $T_e(y,z)=T_{e0}e^{-\alpha z}(1+\beta y)$. Then

\begin{equation}
{\bf B}(z,t)=\frac{cT_{in}}{e}\,(\alpha\beta)\,e^{-\alpha z}(1,0,0)\,t .
\end{equation}
Here $y:0\rightarrow y_m$ with $y_m=R$, while $z:0\rightarrow z_m = h$. If density rises exponentially with $z$, then $\alpha\sim3\times10^{-21}\,$cm$^{-1}$. The temperature profile satisfies $T_e(y_m)=T_e(y_0)e^{-\alpha z}(1+\beta y_m)$ and, for $T_e(y_0)=T_{e0}=10^{3}\,$K, we find  $\beta\sim3\times10^{-20}\,$cm$^{-1}$. These values correspond to characteristic scale lengths $L_\alpha=1/\alpha\sim3\times10^{20}\,$cm and $L_\beta=1/\beta\sim3\times10^{19}\,$cm. 
The ion momentum Eq. (7) yields,
\begin{equation}
{\bf v}_i = - (v_s^2 \beta t e^{- \alpha z}) \hat{y}
\end{equation}
and hence $({\bf v}_i \cdot \nabla)  {\bf v}_i=0 $. The ion velocity is directed along $\hat y$ and depends only on $z$. Consequently, the convective term $({\bf v}_i \cdot \nabla){\bf v}_i$ vanishes identically, and the simplified ion momentum equation, Eq.~(\ref{eq:ionlinear}), remains valid without requiring the timescale restriction given by Eq.~(\ref{eq:taubcondition}). This cancellation is specific to the pressure profile employed in the present astrophysical example. For more general pressure distributions, such as those considered in the laboratory-plasma application discussed below, the convective term need not vanish. In such cases, the validity of Eq.~(\ref{eq:ionlinear}) requires that the magnetic-field generation time $\tau_B$ satisfy the condition given by Eq.~(\ref{eq:taubcondition}).
If we follow Lazarian  \citep{lazarian1992diffusion} and choose $\tau_B=10^9$ $yrs$, then  
Eq. (13) gives $\mid {\bf B}(0,\tau_B)  \mid \simeq 3 \times 10^{-17}$ $G$ and 
$\mid {\bf B}(z_m,\tau_B)  \mid \simeq 10^{-17}$ $G$. 

It is important to note that the ion dynamics not only relate ion flow with seed magnetic field generation, but also contribute in determining the order of magnitude of seed field generation time $\tau_B$. Since pressure gradient is the only source for flow generation, therefore in Eq. (14), we assume the ion fluid velocity at $t=\tau_B$ to be of the order of $v_{sb} = v_s (t=\tau_B)= (T_{out}/m_i)^{1/2}= 9 \times 10^{6}$ $cm/s$. Then Eq. (14) can be used to estimate the seed magnetic field generation time $\tau_B = 3 \times 10^{15}$ $s$ and it yields $\mid {\bf B}(0, \tau_B) \mid = 3 \times 10^{-18}$ $G$ ten times smaller than the previous estimate because our model predicts the seed field generation time ten times smaller than previously assumed time \citep{lazarian1992diffusion}. 
\section{Laboratory Plasma:}
Since in laboratory plasma, the physical quantities and parameters can be estimated more accurately, therefore we apply the present two fluid model to extremely short scale laser-produced plasma. 
The authors \citet{brueckner1974laser} expressed Eq.~(3) as
\begin{equation}
\partial_t{\bf B}=-(cT_e/e)(\nabla n_e/n_e\times\nabla T_e/T_e)
\end{equation}
and assumed
\begin{eqnarray}
\nabla n_e/n_e &=&-\hat{x}|(1/n_{e0})(dn_{e0}/dx)|=-\hat{x}\kappa_n, \nonumber  \\
\nabla T_e/T_e &=&\hat{y}|(1/nT_{e})(dn_{e}/dy)|=\hat{y}\kappa_T
\end{eqnarray}
where $L_n=1/\kappa_n$ and $L_T=1/\kappa_T$ are the characteristic density and temperature gradient scale lengths, respectively. The plasma density and electron temperature were taken as $n_0=10^{20}\,$cm$^{-3}$ and $T_e=10^{3}\,$eV which yields ion sound speed $c_s=(T_e/m_i)^{1/2}\simeq3\times10^{7}\,$cm\,s$^{-1}$. They further assumed $L_n=L_T=0.005\,$cm and the magnetic-field generation time $\tau_B=L_n/c_s\simeq 1.6 \times10^{-10}\,$s. The resulting magnetic field strength was estimated to be $|{\bf B}|\simeq  6 \times10^{5}\,$G.

It may be noted that the ion plasma oscillation time turns out to be
$\tau_i=\omega_{pi}^{-1}=(m_i/4\pi n_0e^2)^{1/2}\simeq8\times10^{-14}\,$s, which is much smaller than the magnetic-field generation time in this plasma. Hence ion dynamics must be considered. In 2023, the plasma flow velocity has been first time measured \cite{tomita2023observation} in a laser plasma experiment and its order of magnitude was $10^{6}$ $cm/s$. In present model, the flow velocity can be determined by using ion dynamics which is coupled with the magnetic field generation mechanism.

To compare our results with the previous work \citet{brueckner1974laser}, we choose $p=p_0(1-\alpha x)$ and $T=T_0e^{\beta y}$ which yield $n=n_0e^{-\beta y}(1-\alpha x)$. Let $x:0\rightarrow x_m$ and $y:0\rightarrow y_m$. Assuming $n(x_m,0)=n_0(1-\alpha x_m)\simeq n_0/3$, we obtain $\alpha=\beta\simeq200\,$cm$^{-1}$ for $x_m=0.003\,$cm and $y_m=0.005\,$cm. We further take $T(0,y_m)=3T_0=3T(0,0)$. Then $L_n=L_T= 5 \times 10^{-3}$ $cm$ and Eq. (3) can be expressed as

\begin{equation}
	{\bf B}(x,y,t)
	=
	\left[
	\frac{cT_{e0}}{e}
	 \frac{(\alpha \beta)
		e^{\beta y}}{(1-\alpha x)} 
	\right]t \hat{z}.
\end{equation}
The ion velocity can be determined using ion equation of motion,

\begin{equation}
	{\bf v}_i(x,y,t)
	=
	v_s(v_s\alpha t) 	\left[
	\frac{
		e^{\beta y}}{(1-\alpha x)} 
	\right] \hat{x}
\end{equation}
where we assume $v_s^2=(1.1)c_s^2$ for $T_i=0.1T_e$. 

In the previous investigation \citep{brueckner1974laser}, the time of generation was assumed to be $\tau_B= \frac{L_n}{c_s}=(1.66) \times 10^{-10}$ $s$. But this value of time does not satisfy the condition (8). Furthermore, $\omega_{pi}^{-1} \ll \tau_B$ in this experiment and ion dynamics should not be ignored. Both ${\bf B}(x,y,t) $ and $	{\bf v}_i(x,y,t)$ are linear functions of time which is unknown. But in experimental plasma we know that the pulse duration time was $t_l \simeq 10^{-9}$ $s$ and hence we should choose $\tau_B \ll \frac{L_n}{v_s} \ll t_l$. Therefore, we assume $\tau_B = 10^{-11}$ $s$ which yields, ${\bf B}(0,0,\tau_B) = 4 \times 10^{4} (0,0,1) $ $G$, ${\bf B}(x_m,y_m,\tau_B) = (2.4) \times 10^{5} (0,0,1)$ $G$, ${\bf v}_i (0,0,\tau_B) = (2.1) \times 10^{6} (1,0,0)$ $cm/s$, ${\bf v}_i (x_m,y_m,\tau_B) = (13) \times 10^{6} (1,0,0)$ $ cm/s$ and condition (8) remains valid. Thus the order of plasma ablation velocity is also determined using the present self-consistent theoretical model.

\section{Conclusion and Future Work}

A self-consistent theoretical model for two fluid plasma evolution has been presented which successfully describes the simultaneous generation of plasma flows and magnetic fields. It has also been pointed out that profiles of plasma density and temperatures are not independent; rather, they are connected through the structural constraint $\nabla^{2}p=0$ on the total pressure profile. This constraint arises from the compatibility between the ion momentum and mass conservation equations in the two-fluid plasma model in the limit $\mid (\mathbf{v}_i \cdot \nabla)\mathbf{v}_i \mid \ll \mid \partial_t {\bf v}_i \mid $.

An important outcome of the present work is the identification of thermal pressure gradients as an efficient mechanism for generating plasma flows. Because the pressure force acts on both electrons and ions, the two species are accelerated in nearly the same direction and with comparable velocities. As a result, the net current remains small, leading to relatively weak magnetic-field generation even when substantial plasma flows are produced. The model therefore explains how strong pressure-driven outflows and weak seed magnetic fields can naturally arise simultaneously from the same physical process.

Interestingly, the density and temperature profiles employed by \citet{lazarian1992diffusion} for estimating seed magnetic fields in galactic gas clouds can be interpreted as a particular realization of the harmonic-pressure solutions obtained in the present work. Within this self-consistent framework, ion dynamics not only determine the associated plasma-flow velocity but also provide an estimate of the magnetic-field generation time $\tau_B$. The theory thus links pressure profiles, flow generation, magnetic-field production, and the characteristic evolution timescale within a single analytical description.

\bibliographystyle{aasjournal}
\bibliography{bibliography}

\end{document}